\documentstyle[11pt]{article}
\newcommand{\be}{\begin{equation}}
\newcommand{\ee}{\end{equation}}

\begin{document}

\begin{titlepage}

\vskip 2. cm
\begin{center}
\vfill
{\large\bf  Why Bohm's Quantum Theory?}
\vskip 3cm

{\bf H. D. Zeh}
\vskip 0.4cm
Institut f\"ur Theoretische Physik, Universit\"at Heidelberg,\\
Philosophenweg 19, D-69120 Heidelberg, Germany.\\
e-mail: zeh@urz.uni-heidelberg.de
\end{center}

\vskip 2 cm
\begin{center}
{\bf Abstract}
\end{center}
\begin{quote}
This is a brief reply to Goldstein's article on ``Quantum
Theory Without Observers'' in Physics Today. It is pointed out
that Bohm's pilot wave theory is successful only because it keeps
Schr\"odinger's (exact) wave mechanics unchanged, while the rest
of it is observationally meaningless and solely based on
classical prejudice.

\vskip .5cm

 \end{quote}

\vskip 1.5 cm
\noindent
\center
To be published in Foundations of Physics Letters

\end{titlepage}

In his recent article on  ``Quantum Theory Without
Observers''
\cite{1}, Sheldon Goldstein raised a number of important
questions about quantum theory. Even though I cannot quite
understand what a universal physical theory without any concept
of observers could {\it mean}, I agree  with many of his critical
remarks. Bell, who once objected ``against measurement'', also
pointed out that Bohm's quantum theory depends on the {\it
assumption} that only its hypothetical classical variables can
directly affect the consciousness of an observer
\cite{2}. However, Goldstein's characterization of the
consequences that have to be drawn from his criticism appears
one-sided, since he neglects many essential aspects.

In particular, he does not mention at all that Bohm's classical
trajectories can neither be experimentally  confirmed nor
refuted if the theory is exactly valid. It is {\it
always} possible to postulate otherwise unobservable
(hence arbitrary) causes for stochastic events.
Bohm's presumed ensemble of classical configurations is merely
 {\it dynamically} consistent because of their presumed
unobservable dynamics. The resulting
trajectories are thus entirely based on classical prejudice.
Goldstein quotes Bell that ``it should be clear from the smallness
of the scintillation on the screen that we have to do with a
particle." Should it then not be as clear that the photon (or any
other boson) is a particle, too, rather than the consequence of a
{\it field} on three-dimensional space, as is assumed in Bohm's
theory? What is observed as a local object is in {\it all} cases
the position of a macroscopic ``pointer" (such as a flash on the
screen or a track in the bubble chamber).

Appropriately localized  {\it wave packets}  representing pointer
positions could be explained as emerging in a fundamental
irreversible ``spontaneous localization'' process by means of a
modified Schr\"odinger equation, such as it has been proposed
by Pearle or Ghirardi, Rimini and Weber (GRW) \cite{3}
and also mentioned by Goldstein. Why then re-introduce classical
variables? There is no need (or any good reason) even for a
Heisenberg picture in terms of {\it particles}, nor for a
wave-particle ``dualism''. And why always represent these
classical variables by a (crucial) ensemble, while the wave
function (the other element of {\it reality} in Bohm's theory) is
normally regarded as given, but tacitly excluded from being
directly experienced by an observer?

The  nonlinear terms in spontaneous localization models are chosen
to become relevant in measurements or similar situations in order
to describe von Neumann's collapse of the wave function. However,
equivalent effects in the density matrix of any relevant {\it
system} arise without any modification of the Schr\"odinger
equation if only the unavoidable environment of this system is
realistically taken into account by what Goldstein calls ``Zurek's
decoherence"
\cite{4,5}. Therefore, these two quite different dynamical
mechanisms (the disappearance and the dislocalization of phase
relations) can both explain the emergence of apparent
``particles" and other quasi-classical (local) properties (such as
Bell's ``small scintillations"). They cannot practically be
distinguished from one another in these cases, provided only the
usual probability rules apply {\it somewhere} and {\it for
something} along the chain of interactions between pointer and
observer(s) in order to justify the concept of a density matrix.
Goldstein's claim that ``the environment acts as an observer" is
quite wrong: decoherence is an objective physical process that is
essential (and unavoidable) also in Bohm's theory for correctly
guiding the postulated classical pointer positions or other
macroscopic variables.

The  application of a  probabilistic collapse
just where environmental decoherence occurs anyway is thus as much
a prejudice as the existence of classical particles and fields.
``Appearences are misleading." Here I fully agree with
Goldstein (although appearance requires an observer). In contrast
to Bohm, Pearle and GRW proposed modifications of quantum theory
that can be experimentally refuted (and very probably have been
unless modified again
\cite{6}).  While the search for deviations from the Schr\"odinger
equation is a reasonable endevour, one must warn young scientists
against spending their time on calculating Bohm trajectories or
``unraveling" entangled (open) systems --- as suggestive as these
pictures may appear to the traditionalistic mind. They would be
investigating mere phantoms. Such calculations would be meaningful
if they led to new observable consequences. However,
precisely this is excluded by construction of Bohm's theory, while
a quantum system that is entangled with its environment clearly
does not possess any wave function by its own.  The idea of
``quantum trajectories" for all systems of interest, now quite
popular in quantum optics, is (1) inconsistent (as it would depend
on the precise choice of the systems), (2) incompatible with a
Schr\"odinger equation (even if that were complemented by a
phenomonological collapse), and (3) in conflict with relevant
experiments.

Most experiments are
local and {\it not} relevant for this purpose (thus permitting
the concept of a local density matrix, or an {\it apparent}
ensemble of wave functions representing it). However, all basic
quantum experiments performed during recent decades have confirmed
consequences of a {\it nonlocal wave function(al)}, defined on a
high-dimensional space that we are used to interpret as a {\it
classical configuration space}. These consequences appear as
``paradoxes''  when described in terms of particles or spatial
waves (fields). The  Schr\"odinger equation,  which unavoidably
leads to drastic (eventually universal)  entanglement between all
macroscopic systems (and thereby {\it locally} to decoherence)
\cite{7},
would not have to be modified at all by means of the apparently
required probalistic terms if one dropped the further prejudice
that there is only one state of each observer in this
``nonlocal reality". A
universally exact Schr\"odinger equation would require that there
is an ever increasing number of different ``versions" of the
quasi-classical world with all its observers, which are
individually described by robust wave packets in appropriate
factor spaces of this high-dimensional space. According to wave
mechanics, observers can only exist in dynamically branching,
thereafter essentially autonomous {\it components} of the global
wave function. These observer versions would practically lack any
possibility of communicating with one another, and therefore need
not be assumed to disappear from reality with precisely one
exception (as postulated by the collapse). Because of their
dynamical autonomy, these global branch wave functions describe
``consistent histories'' to their observers. In order to be
subjectively and individually experienced, a  component of the
global wave function seems to have to possess the product form
$\psi_{obs}\Psi_{rest}$, with a  sufficiently complex and robust
observer state
$\psi_{obs}$, and regardless of the existence of any {\it
classical} observer variables. The observer is the only
``system'' that has to be conceptually defined in principle.

All we have then still to assume in order to
explain Born's probabilities in the form of observable frequencies
of measurement results is that {\it we}, described by certain
correlated
$\psi_{obs}$'s, happen to live in a branch that does not possess a
very small norm. This robust component would represent an
``apparent reality" to us (in the {\it
operational} sense), that also forms the basis of Zurek's
``existential interpretation" \cite{8}.

It is usually
overlooked that Bohm's theory contains the {\it same} ``many
worlds" of dynamically separate branches as the Everett
interpretation (now regarded as ``empty'' wave components), since
it is based on precisely the same (``{\it absolutely} real'')
global wave function
\cite{2}. Its robust components {\it branch} by means of
decoherence (rather than combining by means of recoherence)
because of a fundamental initial condition to the global wave
function that is also responsible for the Second Law
\cite{9} --- not because of ``increasing knowledge'' about a
classical state (the reduction of Bohm's ensemble). Only the
``occupied" wave packet itself is thus meaningful, while the
assumed classical trajectory would merely point at it:
``This is where {\it we} are in the quantum world.'' However, this
can be done without using a trajectory. Any pointing finer
than compatible with the width of the robust wave packet is
empirically unjustified. John Bell (who rejected Everett's
interpretation for being ``extravagant") seems to have realized
this equivalence before he began to favor spontaneous localization
(such as GRW) over Bohm's theory
\cite{10}. This is another (historical) fact simply neglected by
Goldstein when he heavily relies on Bell's words in his
arguments.

% \eject

\end{document}